# Multiple Bound States in the Continuum: Towards Intense Terahertz-Matter Interaction

Quanlong Yang[1, #, *], Zhibo Yao[2, #], Lei Xu[3, #], Yapeng Dou[1], Lingli Ba[1], Fan Huang[2], Quan Xu[2], Longqing Cong[4], Jianqiang Gu[2], Junliang Yang[1,*] Mohsen Rahmani[3], Jiaguang Han[2, *] and Ilya Shadrivov[5]

[1]School of Physics, Central South University, Changsha 410083, China

[2]Center for Terahertz Waves, Tianjin University, Tianjin 300072, China

[3]Advanced Optics and Photonics Laboratory, Department of Engineering, School of Science & Technology, Nottingham Trent University, Nottingham NG11 8NS, UK

[4]Department of Electrical and Electronic Engineering, Southern University of Science and Technology, Shenzhen 51805, China

[5]Nonlinear Physics Centre, Research School of Physics, Australian National University, Canberra ACT 2601, Australia

#, These authors contributed equally to this work.

*Corresponding author. quanlong.yang@csu.edu.cn; junliang.yang@csu.edu.cn; jiaghan@tju.edu.cn;

**Bound states in the continuum (BICs) are an excellent platform enabling highly efficient light-matter interaction in applications for lasing, nonlinear generation, and sensing. However, the current focus in implementing BICs has primarily been on single sharp resonances, limiting the extent of electric field enhancement for multiple resonances. In this study, we conducted experimental demonstrations to showcase how metasurfaces can enable the control of symmetry-broken and Friedrich-Wintgen BICs by leveraging the asymmetry of split resonant rings. This approach allows for the existence of multiple free-control BIC resonances and tailored enhancement of controlling light-matter interactions. We have conducted further experiments to validate the effectiveness and performance of our approach for identification of the distinct fingerprint of α-lactose with high sensitivity using only one single metasurface. These findings present a novel and efficient platform for the development of miniaturized and chip-scale photonics devices with intense light-matter interaction.**

**Introduction**

The confinement or localization of light is of utmost importance in maximizing the interaction between light and matter in microscale and nanoscale photonics devices. Numerous attempts have been made to confine the electromagnetic energy in a small volume using resonant structures, such as localized surface plasmon polaritons[1,2], cavities[3,4] and Fano resonances[5,6]. In recent years, photonic bound states in the continuum(BICs) have emerged as a highly promising solution for achieving intense light-matter interaction. These BICs exhibit high quality factors and strong enhancement of the electric field inside or outside the resonant meta-atoms[7-10], enabling efficient harmonic generation[11-13], fingerprint spectrum acquisition[14-16] and lasing[17]. BIC is a localized mode decoupled from free space embedded in the continuum spectrum of leaky modes. This localization gives rise to an ultrahigh Q factor or infinite lifetime[9]. BICs have demonstrated the ability to be converted into observable quasi-BICs, which possess finite Q factors, by introducing external perturbations[18]. The exploration of BIC resonances has involved various concepts, such as symmetry-protected BICs, Friedrich-Wintgen BICs and others. Of these concepts, symmetry-protected BICs have garnered significant interest due to their simplified design process, as they solely necessitate the presence of broken symmetry[19-23]. Conversely, Friedrich–Wintgen BICs emerge near the crossings of the uncoupled resonances[24,25].

Metasurfaces have become a prevailing platform for achieving light manipulation[26-29], especially for BIC resonances, offering significant advantages such as flexible design, precise control over light-matter interactions, and direct manipulation of the quality factors[18,20,22,30,31]. While metasurface-empowered BICs have successfully demonstrated single resonance capabilities and the ability to enhance near fields at specific frequencies, their applications in photonics often require coverage of multiple operating frequencies[13,14]. For example, in nonlinear photon generation, the focus has typically been on the enhanced field profile at the pump frequency, neglecting the importance of second and third-harmonic frequencies. Furthermore, various molecules have unique characteristic frequency peaks in their fingerprint spectra. To illustrate, α-lactose exhibits four absorption peaks in the range of 0.1-1.6 THz range[32,33]. Previous approaches have utilized arrays of metasurfaces to cover the entire frequency range of interest, which requires significant fabrication complexity and experimental efforts. Therefore, there is an urgent need for a new methodology or regime that enables the generation of multiple free-control BIC resonances. However, experimental demonstrations in this area are currently lacking.

Here, we propose a novel regime for achieving multiple symmetry-protected and Friedrich-Wintgen BICs at terahertz frequencies. We demonstrate the viability of this approach experimentally using a dual-split-resonant-rings (DSRRs) metasurface. By leveraging the asymmetry of the SRRs, we are able to directly control the positions and linewidths of these BICs, without requiring computationally intensive calculations or strict fabrication accuracy. To analyze the mechanism behind these multiple BICs, we use coupled mode theory and multiple decomposition techniques. Our terahertz time-domain spectroscopy setup allows us to characterize angle-resolved multiple BIC resonances and fingerprint retrieval of $\alpha$-lactose. Compared to conventional single BIC resonances, our designed metasurface-empowered multiple BICs offer additional degrees of freedom for boosting light-matter interactions, making them an ideal platform for nonlinear harmonic generation and molecular fingerprint retrieval. Our study offers a promising pathway for the development of a practical and efficient platform for these advanced applications.

**Metasurface design**

Figure 1a illustrates the schematic of our DSRR metasurfaces for achieving multiple BICs, consisting of a dual SRRs periodic array with a uniform ring width and gap size. For clarity, we designate these two resonators as the inner ring and outer ring. The unit cells are situated on a low-loss and low-index substrate with cyclic olefin copolymer (COC) with $\varepsilon=2.32+0.0147i$ at 1THz to mitigate additional losses. We numerically optimized the DSRR metasurfaces to simultaneously achieve multiple free-control BICs resonance at the interest frequencies. Upon excitation with a THz pulse, a sequence of current oscillations occurs on the rings, leading to the generation of electric dipoles, magnetic dipoles, and higher-order multipoles. The coupling between the inner ring and outer ring results in the emergence of newborn Friedrich-Wintgen BICs. Interestingly, as compared with the conventional Friedrich-Wintgen BICs, our design is directly controlled by the asymmetry of the split resonant rings that is defined by the angle $\alpha$. Both symmetry-protected BICs and Friedrich-Wintgen BICs exhibit remarkably magnified field behaviors in the gaps of the rings, as demonstrated in Fig. 1b and 1c.

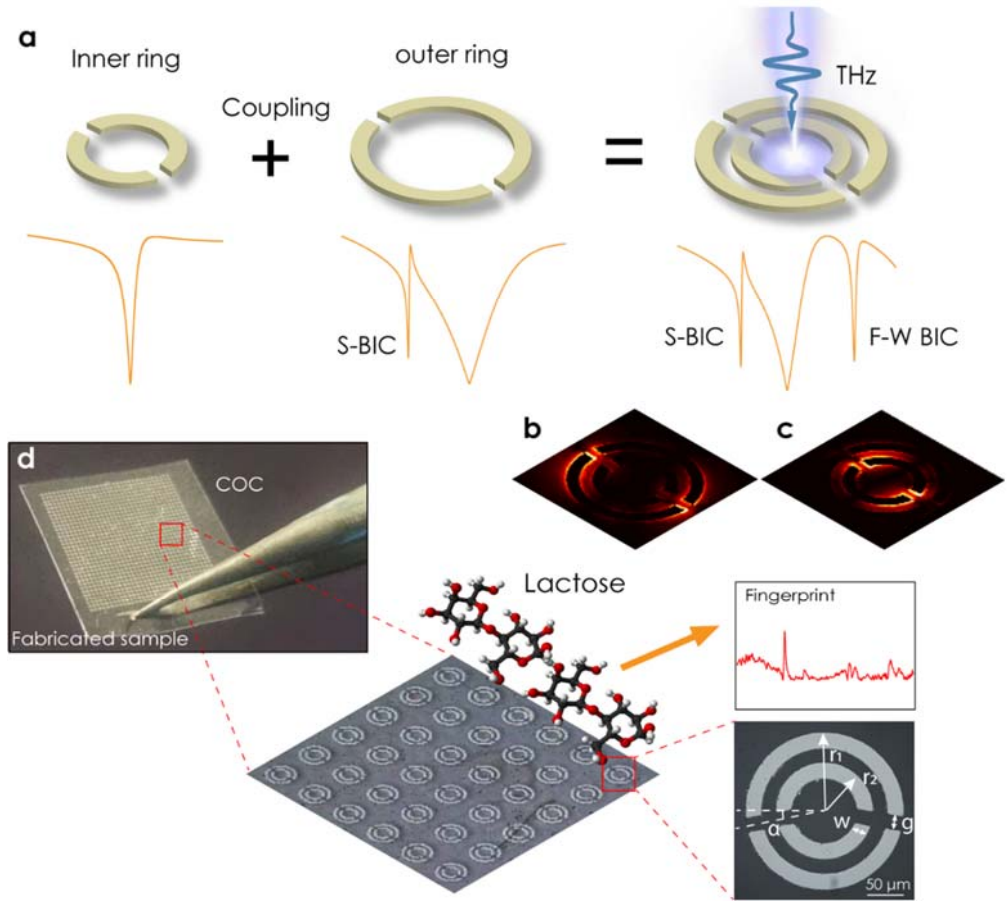

**Figure 1| Multiple BICs in dual split resonator rings**. (**a**) Schematic of the multiple BICs unit cell consisting of dual split resonant rings (Inner ring and outer ring). The below transmission spectrum excited by terahertz waves shows the emergence of newborn Friedrich-Wintgen BICs resulting from the strong coupling between the inner and outer rings. (**b, c**) The electromagnetic hotspot of symmetry-protected and Friedrich–Wintgen BICs in **a**. (**d**) The fabricated multiple BICs metasurface and scanning electron microscopy image of partial samples. The geometrical unit cell parameters are fixed as periodicity $P$ = 210 μm; radii of the outer and inner ring are $r_1$= 80 μm and $r_2$= 50 μm, respectively; width of the ring and gap is $w$ = 15 μm and $g$ = 15 μm, respectively; gaps of outer and inner ring gaps oriented at the same angle. The positioning of the three designed multiple BIC resonances is strategically aligned with the absorption peaks of the coated lactose, thereby ensuring enhanced fingerprint retrieval.

Notably, owing to the excited multipolar content of the modes, the designed multiple BICs can be freely engineered separately through the geometric parameters, such as radius of inner and outer rings, the orientation angles of the inner and outer rings. Such a multiple-BICs platform ensures a robust and efficient approach for enhancing light-matter interactions, as they simultaneously offer strong confined fields at both the pump and target frequencies or diverse characteristic frequency peaks. To demonstrate the advantage of our multiple BICs metasurface for molecular fingerprint retrieval, we present the concept in Fig. 1d. By carefully designing multiple BICs to encompass all these absorption frequencies of α-Lactose, we coated α-Lactose onto the metasurface, allowing the molecules to distribute freely and attach to the gaps of the rings. The confined electric field generated by multiple BICs significantly enhances the performance of fingerprint retrieval at the multiple absorption frequencies, see the inset of Fig. 1d. In comparison to traditional single BICs, our design enables the measurement of multiple peaks simultaneously with a single metasurface, which previously required multiple metasurfaces.

**Numerical simulations**

To demonstrate the spectral response of our proposed DSRR metasurface, we conducted a spectrum analysis using a finite-element numerical calculation method and extracted the field profiles of DSRRs, which are shown in Figure 2. Notably, BIC resonances can be excited under both *x*- and *y*-polarized light, our focus lies on the horizontal polarization (*x*-polarization) parallel to the direction of the gaps ($\alpha=0°$). Figures 2a and 2b illustrate the transmission spectra of the inner ring and outer ring via changing the orientation angle of the gaps. For the inner ring, a quasi-BIC resonance arises around 0.8 THz due to the breaking of symmetry and the radiative Q-factors show $\alpha^{-2}$ dependence, while for the outer ring, another symmetry-protected BIC is triggered at 0.46 THz. The Q factors of these two BICs drop from infinite to finite as the orientation angle increases. Additionally, two resonances with large linewidth emerge at 0.6 THz and 1.0 THz for the outer and inner rings, respectively. It worth noting that the spectral positions of these two BICs can be separately controlled through the geometry size of the inner and outer rings.

After the combination of the inner and outer rings, a significant change becomes evident due to the interplay between the inner and outer rings, as depicted in Figure 2c. To clarify the underlying mechanism governing the emergence of multiple BICs, we systematically partition the transmission of the metasurface into three regions, namely Region I, II, and III. The symmetry-protected BIC denoted as M1, located at a frequency of 0.46 THz, corroborates our prior findings established for the outer ring as the gap orientation angle increases. In contrast, the behaviors exhibited by M2 and M3 manifest the distinctive 'avoided crossing' behavior, reminiscent of the Friedrich-Wintgen BIC scenario. Specifically, when α is set to 0°, the symmetry-protected BIC denoted as M3 persists in its observability as the case of the inner ring. However, a discernible decline in the Q factors of M3 becomes apparent as the angular orientation gradually increases. Intriguingly, with the imposition of in-plane symmetry breaking, the linewidth associated with M2 demonstrates a steep reduction, a phenomenon markedly distinct from the behavior observed in the outer ring counterpart. Upon reaching an angular orientation of α = 28°, a convergence between the behaviors of M2 and M3 is observed, leading to the occurrence of an avoided crossing. This intriguing phenomenon is accompanied by the vanishing linewidth of M2 and the attainment of the minimal quality factor for M3. Continuing along this trajectory, as α undergoes further augmentation, an inversion in the previously observed trends for M2 and M3 comes to the fore. Analogously, the resonances labeled as M4 and M5 follow analogous patterns characterized by avoided crossings, thereby underscoring the shared underlying physical mechanisms.

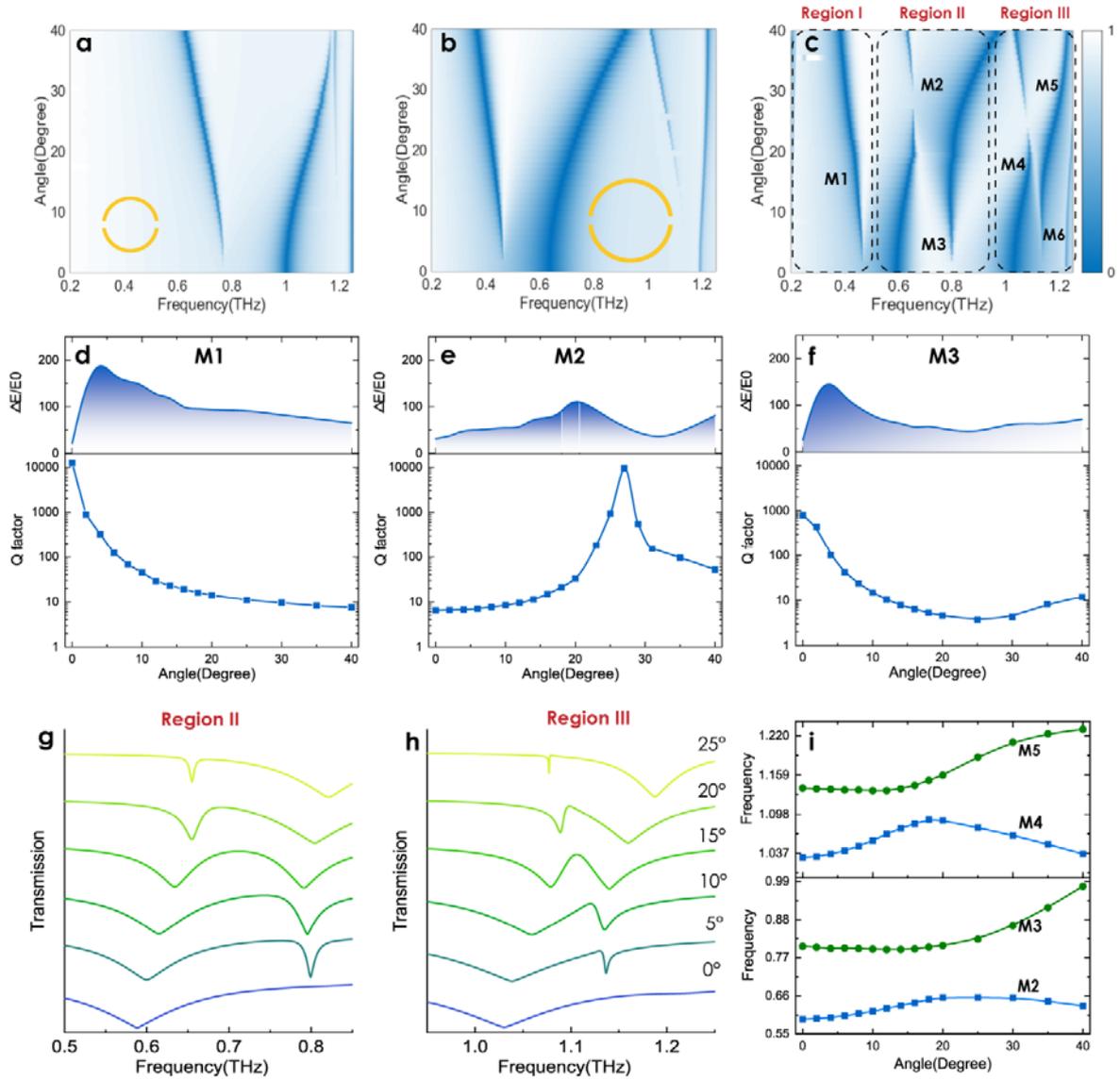

**Figure 2| Evolution and performance of multiple BICs. (a, b,c)** Transmission spectra of inner ring, outer ring, and multiple BIC metasurface via changing the orientation angle of the gap. The transmission spectra of multiple BICs metasurface are categorized into three parts based on the underlying evolution mechanism, Region I, Region II, and Region III. All resonances are marked as M1- M6. **(d, e, f)** Electric field enhancements and quality factors of M1, M2, and M3 for different orientation angles. The maximum field enhancement is achieved when both intrinsic and radiative damping rates of the BIC mode are matched. The maximum field enhancement of M1-M3 reached the value of 202,126.4 and 154, respectively. **(g,h)** Transmission fitting curves of Regions II and III w were obtained using the coupling mode theory versus the orientation angle. **(i)** Frequency dispersion fitting of both Region II and III. The solid line represents the simulation frequency of M2 to M5, and the inverted triangle denotes the corresponding fitting frequencies.

To explore the evolutionary trends of multiple BICs, we conducted an analysis that involved extracting data related to electric field enhancements and quality factors of three specific modes: M1, M2, and M3. These values were obtained through eigenfrequency analysis and are represented in Fig. 2d-f. For BIC modes M1, the radiative Q factors followed a distinct $\alpha^{-2}$ dependence characteristic of symmetry-protected BICs versus the orientation angle. Interestingly, the highest field enhancement of M1 did not coincide with the point of highest radiative Q factors, as depicted in Fig. 2d. This discrepancy arises because the overall quality factors of the BICs are decomposed into radiative and intrinsic components. The intrinsic aspect encompasses non-radiative losses originating from material properties and structural morphology. The highest field enhancement was achieved when a balance was matched between the radiative and intrinsic Q factors[32]. In Fig. 2e and 2f, the evolution of M2 and M3 demonstrated an opposing trend due to their mutual coupling dynamics. In the case of M2, the radiative Q factor initially increased and subsequently decreased, with a peak occurring at the point of α = 28°. In addition, the Q factor of M3 exhibited its lowest value at the same angle as well. Similarly, for both M2 and M3, the highest field enhancement was not achieved at the maximum radiative Q factor. It is worth highlighting that the field enhancement for M1 and M3 reached its summit at approximately α = 5°, which is an ideal parameter for optimizing metasurface performance. The corresponding analysis of M4 and M5 could be seen from the supporting information.

We further employed the temporal coupled mode theory to analyze the coupling behavior of BICs in Regions II and III. The transmission of the coupled systems could be described as follows[34,35]:

$$T = t_0 \frac{(\omega_1 - \omega + i\gamma_{i1})(\omega_2 - \omega + i\gamma_{i2}) - \beta^2}{(\omega_1 - \omega + i\gamma_{i1} + i\gamma_{e1})(\omega_2 - \omega + i\gamma_{i2} + i\gamma_{e2}) + \left(p\sqrt{i\gamma_{e1}i\gamma_{e2}} - i\beta^2\right)}, \quad (1)$$

*Here, $\omega_1$, $\gamma_{i1}$ and $\gamma_{e1}$ ( $\omega_2$, $\gamma_{i2}$ and $\gamma_{e2}$) are the frequency, intrinsic loss and radiative loss of the modes, respectively. $p$ denotes the phase difference and $\beta$ represents the coupling parameters between the two modes. Figures 2g and 2h display the transmission fitting curves of the orientation angle change from 0° to 25°. The fitting results exhibit a good agreement with the simulated results. We note that the coupling strength of two modes gives rise to the changes in the line width of multiple BICs, and the coupling strength could be determined by the asymmetry of dual SRRs. The frequency dispersion of Regions II and III from the coupled mode theory (dots) and simulated results (solid line) are shown in Fig. 2i. The smallest spectral splitting of Region II is 0.15 THz, occurring at $\alpha$ = 20°, while Region III exhibits a splitting of 0.06 THz at an angle of 16°. The disparity in coupling strengths among the multipoles introduces variations in the spectral splitting between regions II and III.

**Multipole analysis of multiple BICs**

The properties of these modes can be described through the multipoles analysis. Each mode can be viewed as a superposition of electric and magnetic multipoles. Modes sharing the same multipole contents will interfere with each other when their spectral positions are closely aligned, leading to an exchange and variation in their multipolar contents. This will further result in a suppression or increase on their radiation accordingly. To analyze the multipole nature of multiple BICs, we start by considering the multipoles of the metasurface via spherical multipole decomposition. The total scattering cross-section of the metasurface is given as [36]:

$$C_S = \frac{\pi}{k^2} \sum_{l=1}^{\infty} \sum_{m=-l}^{l} (2l+1)\left[\left|\alpha_E(l,m)\right|^2 + \left|\alpha_M(l,m)\right|^2\right], \quad (2)$$

where $\alpha_E$ and $\alpha_M$ are the coefficients of the electric and magnetic multipoles, respectively; ($l$, $m$) is the order of the electric and magnetic multipoles. The multipolar excitation strength of each multipole is calculated by considering the fields radiated into the air and obtaining its contribution to the total scattering cross-section as described in Eq. (2).

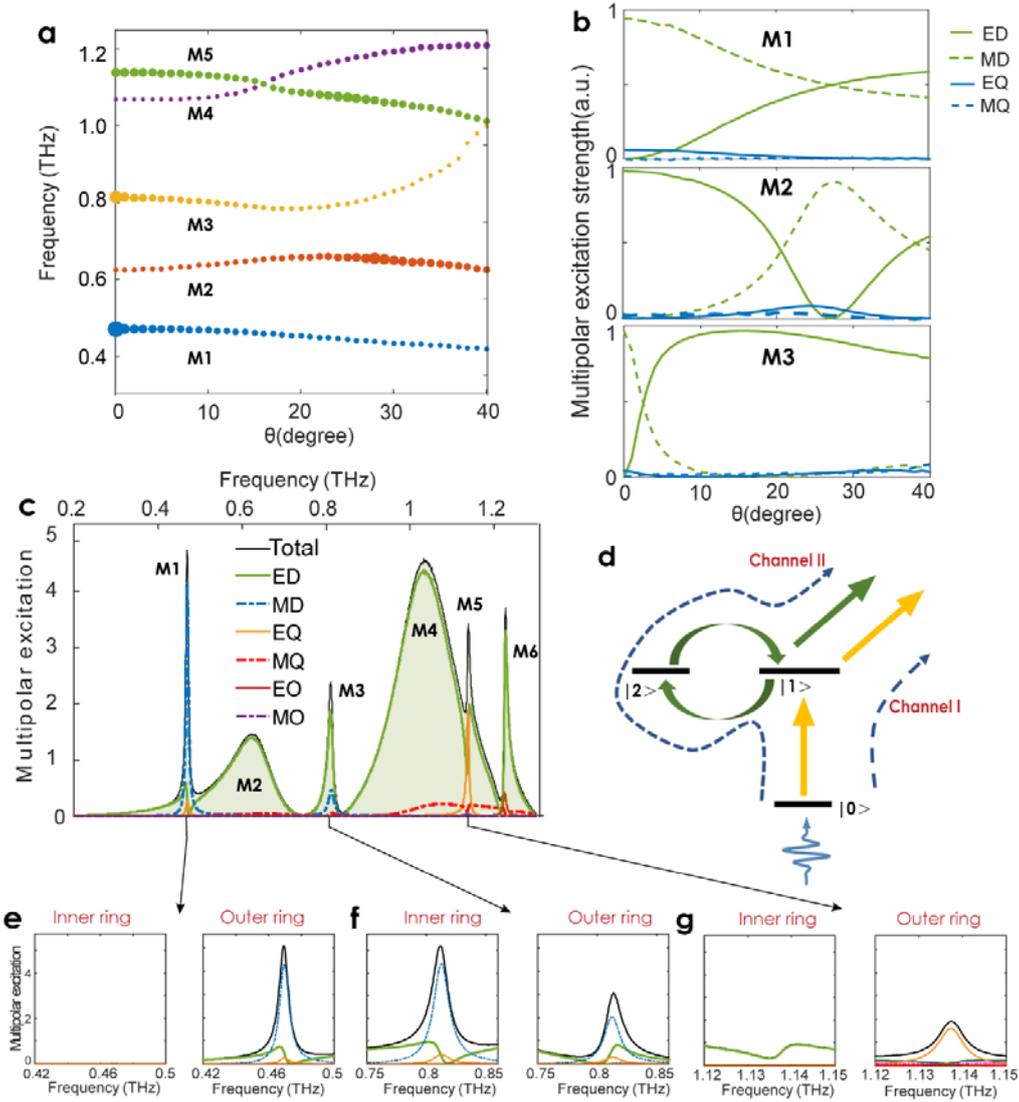

**Figure 3| Multipoles analysis of multiple BICs. (a)** Dispersion diagram of modes M1 – M5 with varying $\alpha$. **(b)** Multipolar structures for modes M1, M2, and M3 with varying $\alpha$. **(c)** Calculated scattering cross-section of Total, electric dipole, magnetic dipole, electric quadrupole, magnetic quadrupole, electric octupole and magnetic octupole from metasurface with $\alpha = 5°$. **(d)** Coupling regimes of multiple BICs. The ground state |0> represents the *x*-polarization wave pump, while |1> denotes the multipoles acting as the bright mode, |2> and |3>

correspond to the dark modes. Solid and dashed arrows represent the energy flow and leaky channel. **(e, f, g)** Calculated contributions to scattering cross-section from inner and outer rings for modes M1, M3, and M5.

Figure 3a gives the dispersion diagram for the eigenmodes of M1–M5 with varying orientation angle theta. The size of the dot indicates the logscale Q factors of the modes, which has also been quantitatively shown in Figure 2b. The dispersion diagram of the eigenmodes M1–M5 matches well with the transmission spectra in Figure 2c, indicating the direct access to these modes under *x*-polarized pump incidence from the free space. Figure 3b gives the multipolar transformation for modes M1, M2, and M3, respectively. Take mode M1 as an example, as can be seen, for $\alpha$ = 0°, a symmetry-protected BIC magnetic dipole (MD) is formed at the Γ point of the metasurface system[12]. With increasing α, a leaky channel - electric dipole (ED) excitation is gradually opened, and the Q-factor of M1 decreases. For M2 and M3, as shown in Figure 3a, owing to the coupling between these two adjacent modes when their spectral positions approach each other, we observe typical anti-crossing features from these mode pairs, indicating the formation of FW BIC[37]. From the multipolar analysis shown in Figure 3b, the interference between M2 and M3 leads to a decrease of the leaky channel ED in the multipolar content of M2 around α = 28°, resulting in the formation of FW-type BIC for M2, manifesting a peak in the Q-factor curve (see Figure 3a and Figure 2e). In contrast, with increasing *α*, the nonradiative channel MD in mode M3 gradually transforms to M2 and leads to a decrease in its Q-factor as illustrated in Figure 3a and Figure 2f. Similar multipole model and multipolar transformation can be observed for modes M4 and M5, for detailed information see the Supporting Information.

In the following, we consider $\alpha$ = 5° as an example and examine the multipolar excitation by *x*-polarized light. Figure 3c illustrates the calculated total scattering cross-section and contributions from the multipoles. It is worth noting that different multipoles play significant roles in the performance of BICs. Specifically, for M1, the optical response is primarily dominated by the MD, with smaller contributions from the ED. To further demonstrate the regime of BICs, we calculated the scattering cross-section of the inner and outer ring, as shown in Figs. 3e-g. Additionally, we extracted the surface current and electric field distribution of the metasurface, which could be seen from the supporting information. Here, an out-of-plane MD emerges from two outer rings, exhibiting zero overlap with the only radiative wave in such a subdiffraction-limited system, indicating this MD is completely bounded. However, by changing the angle α to break the symmetry of the dual SRRs, a leaky channel (ED) near the gap opens up when illuminated with an *x*-polarized wave pump. Figure 3d displays the coupling mechanism of BICs M1, where bright state|1> represents ED and out-of-plane MD corresponds to dark state |2>. This broken symmetry facilitates the interaction between the pump wave and the EDs, enabling the external excitation to couple into the out-of-plane MD through the ED. Interestingly, the Fano shape is consistently associated with the excitation of the leaky channel (ED), as it represents the channel responsible for energy exchange between the subdiffraction-limited system and the far-field radiation. With the increment of angle $\alpha$, the overlap between the external pump and MD becomes stronger, and thus the coupling broadens the line width and reduces Q-factors of quasi-BIC. The corresponding analysis of modes M2 and M4 could be seen from the supporting information.

Next, we applied the same methodology to demonstrate the performance of the quasi-BIC mode M3, it is evident that the total optical response is predominantly influenced by ED and MD contributions. The coupling mechanism observed in Figure 3d for M1 is also applicable to BICs M3. As shown in Fig. 3f, two bounded MDs are formed by contributions from inner and outer rings, respectively. Similarly, the introduction of broken symmetry leads to the emergence of a leaky channel ED from both the inner and outer rings. Due to the out-of-phase nature of the MD modes from the inner and outer rings, the total scattering cross-section contribution from MD in the metasurface is smaller than that of a single ring. In contrast, for the quasi-BIC mode M5, the electrical quadrupole (EQ) plays a prominent role in the scattering cross-section, with ED contributing to a smaller extent. As depicted in Figure 3g, the out-of-plane EQ arises from the outer ring. Unlike modes M1 and M3, the contribution of ED originates from two sources: one is located at the inner ring and can be directly excited by the x-polarized plane wave, while the other is formed between the inner and outer rings due to their coupling. With the introduction of broken symmetry, the ED mode from the inner ring thus becomes the energy exchange channel between the subdiffraction-limited system and far-field radiation. Higher-order BICs such as M6 exhibit a more complex coupling regime, as shown in the supplementary information.

**Experimental demonstrations**

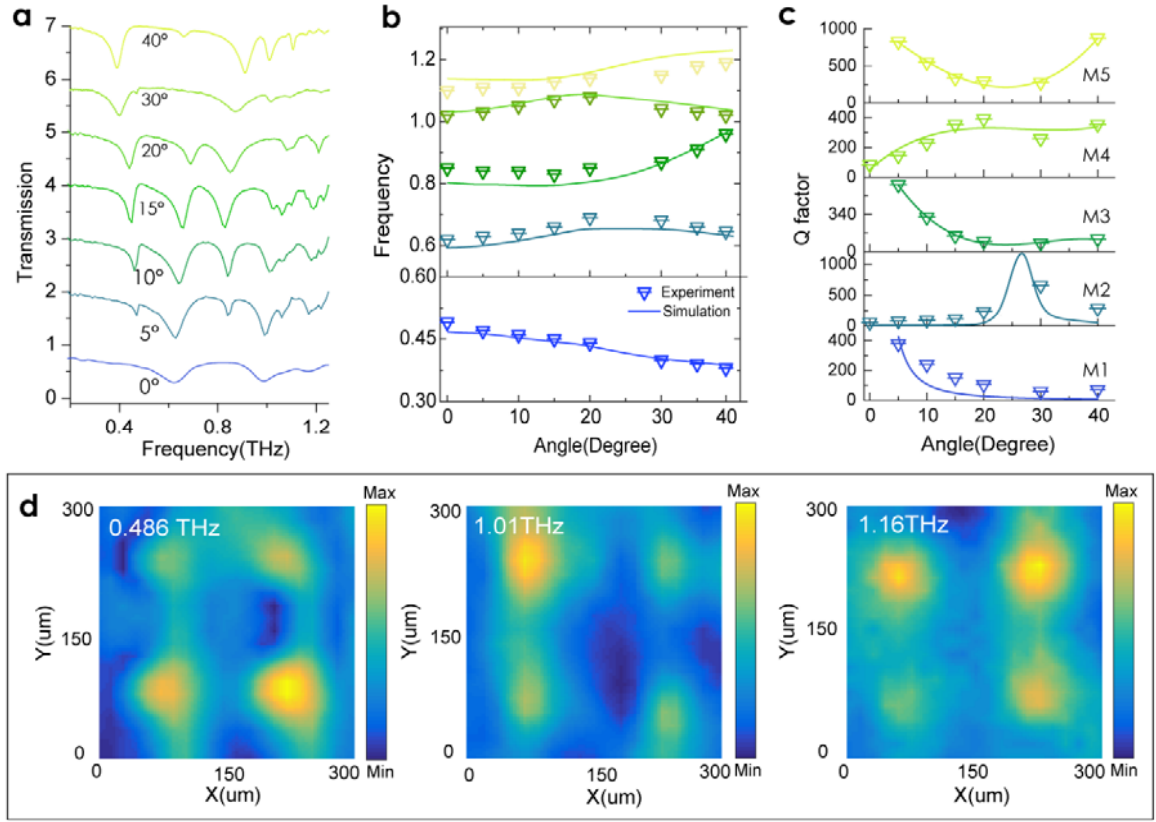

**Figure 4| Experimental demonstration of multiple BICs. (a)** Measured transmission performance of metasurface acquired by varying the orientation angle from 0° to 40°. **(b)** The frequency comparison of experiment and simulation results from the metasurfaces. The solid line represents the simulation frequency of M1 to M5, and the inverted triangle denotes the corresponding measured frequencies. **(c)** Measured Q factors and fitting line of Multiple BICs. The solid line represents the fitting line, and the inverted triangle denotes the corresponding measured Q factors. **(d)** Measured near-field electric field distribution of DSRR metasurface with $\alpha$ = 15°. The hotspot denotes the magnified field distribution in the gaps of the rings.

To experimentally demonstrate the transmission signature of multiple BICs, we fabricated the metasurfaces starting with the commercial COC layer and using photolithography techniques (see Methods for fabrication details). The optical image and partially enlarged image of the fabrication sample can be seen in Fig. 1d. Here, a home-built terahertz time-domain spectroscopy(TDS) and terahertz near-field scanning spectroscopy were used to characterize the performance of metasurface(see Methods for experiment details). Figure 4a displays the measured transmission performance of fabricated samples and it is in good agreement with the simulated results. M1 holds the behaviors of symmetry-protected BIC versus the orientation angle. M2 and M3 (M4 and M5) also follow analogous patterns characterized by avoided crossings as the calculated results. Limited by the time-domain scanning range of TDS and intrinsic loss of Aluminum, some quasi-BICs at higher frequencies are not clearly visible. As shown in Fig. 4b, the measured resonance frequencies of M1 to M5 (inverted triangle) agree well with those from the simulation (solid line), and the small deviation is attributed to the fabrication errors and measured resolution.

Furthermore, we analyzed the behaviors of measured Q factors. We applied Fano fitting to quantify the Q factor of multiple BICs and the results are presented in Fig. 4c. It is evident that the measured Q factors of multiple BICs are consistent with the simulated results in Fig. 2. As the orientation angle increases, the Q factors of M1 decrease in a manner that resembles an inverse quadratic relationship, which aligns with the findings related to symmetry-protected BICs. Interestingly, the Q factor of M2 initially increases from a value of 55 but then decreases after reaching a critical coupling point with the value of 658. In contrast, the Q factor of M3 shows the opposite trend (608 to 78), providing further evidence of the avoided crossing phenomenon between M2 and M3. Additionally, M4 and M5 exhibit similar trends as M2 and M3, as can also be observed in Figure 4c. Furthermore, we measured the electric field distribution of the DSRR metasurface with $\alpha = 15°$ using a terahertz near-field setup, which is shown in Figure 4d. An obvious magnified field distribution could be seen in the gaps of the rings for multiple BICs at 0.486 (M1), 1.01(M3) and 1.16 THz (M5).

**Multiple BIC for THz-matter interactions**

To harness the potential of multiple BICs towards intense light-matter interaction, we have implemented the metasurface to enhance the performance of molecular fingerprint retrieval at terahertz applications. The previous study predominantly focused on low-Q factor resonant structures or individual quasi-BICs, where the limited field enhancement and working frequency range constrained the performance of terahertz sensing. Notably, lactose is a disaccharide composed of D-glucose and D-galactose linked by a β-1,4 glycosidic bond. Given that $\alpha$-lactose exhibits multiple characteristic absorptions at terahertz frequency, our multiple BICs perfectly align with the precise retrieval of α-lactose's fingerprint. Therefore, we undertook a redesign of the metasurface, repositioning BICs M1, M3, and M5 at frequencies of 0.5, 0.95, and 1.17 THz, respectively. To comprehensively evaluate the impact of the Q factors, we introduced three metasurfaces with the orientation angles 5°, 10° and 15°, each with distinct geometrical dimensions to ensure that all three quasi-BICs covered the absorption peaks of lactose. The corresponding simulation results of these metasurfaces are presented with solid lines in Figs. 5a. In the experiments, prior to introducing lactose to metasurfaces, we conducted measurements of the transmission profiles of the metasurfaces using a TDS system. All measured transmission results agree well with the simulation data, as depicted by the green dots in Fig. 5a. There was a minor amplitude mismatch due to potential errors introduced during both the measurement and fabrication processes.

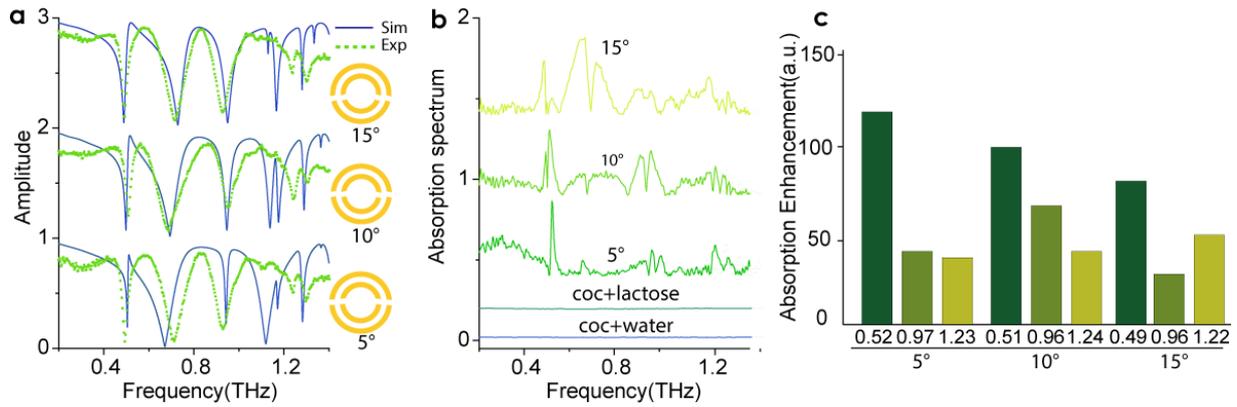

**Figure 5| THz-matter interaction of Multiple BICs. (a)** Simulated (dark blue line ) and measured (green dots) transmission of metasurfaces for lactose sensing. The orientation angles of DSRR $\alpha$ = 5°, 10° and 15°, periodicity $P$ = 210 μm; radii of the outer and inner ring are $r1$= 80 μm and $r2$= 50 μm, respectively; width of ring and gap is $w$ = 15 μm and $g$ = 15 μm . **(b)** Extracted fingerprint spectrum of metasurface with lactose. All the samples are measured with a lactose concentration of 2.77 nmol/μl. The absorption spectrum of the COC layer with lactose and water is also measured as a comparison. **(c)** The absorption enhancement of metasurface with lactose compared to the COC layer with lactose. The maximum absorption enhancement of $\alpha$ = 5°, 10° and 15° reached the value of 116.5, 97 and 78.5, respectively.

Subsequently, α-Lactose was deposited onto the metasurfaces (see Methods for the details), and the dried lactose molecules were uniformly distributed across both unit cells and vacant areas. We applied the concentrations of 2.77nmol/μl to investigate the light-matter interactions of multiple BICs, the measured absorption spectrum is depicted in Fig. 5b. We found that three absorption peaks around 0.52, 0.96 and 1.23 THz arise from the spectrum of both α = 5°, 10°, and 15°. Compared with the performance of a single BIC, multiple BICs realized the fingerprint retrieval of α-Lactose using a single metasurface, which demonstrates the priority of multiple BICs for light-matter interactions. Limited by the time-domain range and fabrication errors, the absorption peaks of the metasurface with different orientation angles display a frequency shift of less than 0.03 THz. As a comparison, the absorption spectrum of the COC layer (without metasurface) with lactose and water are also measured in the same preparation and measurement way. Interestingly, no absorption peaks could be identified from the spectrum of both lactose and water. For a quantitative analysis of multiple BICs, we extracted the data of absorption enhancement, as shown in Fig. 5c, wherein absorption enhancement was determined by dividing the maximum absorption of the lactose-coated metasurface by the reference value of the COC layer with lactose. Notably, the metasurfaces exhibited a distinct absorption enhancement at all absorption peaks of lactose, and the best absorption enhancement reaches 116.5. Another intriguing observation was that the absorption enhancement decreased with increasing orientation angle of the gaps, consistent with experimental expectations of field enhancement.

**Conclusion**

In conclusion, we have developed a platform for generating multiple BICs using a DSRR metasurface. By breaking the symmetry of the SRRs, we can induce both symmetry-broken and Friedrich-Wintgen BICs, and precisely control their resonance positions and linewidths through variation of the SRR orientation angles. Through analysis of the multipoles and coupling regimes, we are able to gain insight into the nature of these BICs, while transmission and quality factor measurements confirm their effectiveness. In addition to their strong sensing capabilities, as demonstrated by the identification of the distinct fingerprint of α-lactose using a single metasurface, these multiple BICs offer a promising platform for intense light-matter interactions and novel terahertz device applications.

**Methods**

**Simulation:** The simulated eigenfrequency and Q factor of multiple BICs are calculated by the eigenmode analysis of COMSOL Multiphysics, where periodic boundaries are used in the simulations. In addition, CST Microwave Studio (the frequency-domain solver with unit cell conditions) is employed to characterize the transmission spectra and field distributions of dual SRRs.

**Fabrication:** We fabricate dual SRRs on a commercial 50 μm COC film with low loss tangent. Initially, the COC layer was attached to a 1 mm-thick silicon wafer to serve as a rigid support.

Standard photolithography is employed to delineate the pattern of the metasurface with the photoresist film (PR4000) and developing solution (RZX3038). Afterward, a 200 nm aluminum layer was thermally evaporated onto the upper surface of the COC and the Al layer was selectively removed using an acetone solution, resulting in the formation of the desired dual SRRs. After the acetone removal and surface cleaning, the free-standing metasurface was removed from the silicon wafer.

**Characterization:** In the experiment, we characterized the fabricated sample using photoconductive-antenna-based THz time-domain spectroscopy (THz-TDS) under normal incidence. To ensure high polarization directivity of the incident wave, we placed an *x*-direction polarizer at the front of the sample, the other settings of TDS systems are chosen as follows: the delay line length is 170 ps, the accuracy is 0.02 ps, the average scanning times is 3, and the chamber was purged with dry nitrogen to avoid noises from water vapor absorption. For the preparation of α-Lactose, a lactose dilute solution with a concentration of 2.77nmol/uL was prepared by dissolving α-lactose monohydrate in deionized water. Firstly, a pipette gun was used to drop 20 μL lactose aqueous solution into the center region of metasurfaces, and the solution was placed on a hot plate at 80℃ and heated at constant temperature for about 8~10 minutes. After the droplets were dried, the metasurfaces with the α-lactose were characterized with the same THz-TDS. For the near-field measurement of the metasurface, the electric field component $E_x$ was detected by a near-field scanning terahertz microscopy system. A fiber-coupled terahertz near-field probe with a resolution of up to 10 μm was used as the detector was mounted on a two-dimensional translation stage which enabled 2D scans at a fixed distance from the sample surface. The 2D electric field was detected with a 20 μm step in the *x* direction and *y* direction from −0.15 mm to +0.15 mm.

## Acknowledgments

The authors thank Dr. Mingkai Liu and Dr. Dongyang Wang for helpful discussions. The authors would like to acknowledge the financial support from the National Natural Science Foundation of China( Grant No. 62205380).

## Author contribution

Q.Y., L.X.,J.Y., J.H. and I.S. developed the idea and designed the multiple BIC metasurfaces. Q.Y., Y. D., L.B. and L.X. conducted numerical simulations. Q.Y and Z.B. fabricated the samples, Q.Y., Z.B., and Q.X. performed the experimental measurements, Q.Y., L.X., J.Y., J.H. and I.S. analyzed the data and wrote the manuscript. All authors discussed the results and contributed to the manuscript writing.

**Disclosures.**

The authors declare no conflicts of interest.

**Reference**

1  Meinzer, N., Barnes, W. L. & Hooper, I. R. Plasmonic meta-atoms and metasurfaces. *Nat. Photon.* **8**, 889-898, (2014).
2  Bin-Alam, M. S. *et al.* Ultra-high-Q resonances in plasmonic metasurfaces. *Nat.Comm.* **12**, 974, (2021).
3  Yang, S., Wang, Y. & Sun, H. Advances and Prospects for Whispering Gallery Mode Microcavities. *Adv. Opt. Mater.* **3**, 1136-1162, (2015).
4  Yu, D. *et al.* Whispering-gallery-mode sensors for biological and physical sensing. *Nat. Rev. Methods Primers* **1**, 83, (2021).
5  Srivastava, Y. K. *et al.* Ultrahigh-Q Fano Resonances in Terahertz Metasurfaces: Strong Influence of Metallic Conductivity at Extremely Low Asymmetry. *Adv. Opt. Mater.* **4**, 457-463, (2016).
6  Limonov, M. F., Rybin, M. V., Poddubny, A. N. & Kivshar, Y. S. Fano resonances in photonics. *Nat. Photon.* **11**, 543-554, (2017).
7  Hsu, C. W. *et al.* Observation of trapped light within the radiation continuum. *Nature* **499**, 188-191, (2013).
8  Zhen, B., Hsu, C. W., Lu, L., Stone, A. D. & Soljačić, M. Topological Nature of Optical Bound States in the Continuum. *Phys. Rev. Lett.* **113**, 257401, (2014).
9  Hsu, C. W., Zhen, B., Stone, A. D., Joannopoulos, J. D. & Soljačić, M. Bound states in the continuum. *Nat. Rev. Mater.* **1**, 16048, (2016).
10 Koshelev, K., Bogdanov, A. & Kivshar, Y. Meta-optics and bound states in the continuum. *Sci. Bull.* **64**, 836-842, (2019).
11 Carletti, L., Koshelev, K., De Angelis, C. & Kivshar, Y. Giant Nonlinear Response at the Nanoscale Driven by Bound States in the Continuum. *Phys. Rev. Lett.* **121**, 033903, (2018).
12 Xu, L. *et al.* Dynamic Nonlinear Image Tuning through Magnetic Dipole Quasi-BIC Ultrathin Resonators. *Adv. Sci.* **6**, 1802119, (2019).
13 Koshelev, K. *et al.* Subwavelength dielectric resonators for nonlinear nanophotonics. *Science* **367**, 288-292, (2020).
14 Tittl, A. *et al.* Imaging-based molecular barcoding with pixelated dielectric metasurfaces. *Science* **360**, 1105-1109, (2018).
15 Leitis, A. *et al.* Angle-multiplexed all-dielectric metasurfaces for broadband molecular fingerprint retrieval. *Sci. Adv.* **5**, eaaw2871, (2019).

16   Liu, B. *et al.* Terahertz ultrasensitive biosensor based on wide-area and intense light-matter interaction supported by QBIC. *Chem. Eng. J* **462**, 142347, (2023).
17   Kodigala, A. *et al.* Lasing action from photonic bound states in continuum. *Nature* **541**, 196-199, (2017).
18   Koshelev, K., Lepeshov, S., Liu, M., Bogdanov, A. & Kivshar, Y. Asymmetric Metasurfaces with High-Q Resonances Governed by Bound States in the Continuum. *Phys. Rev. Lett.* **121**, 193903, (2018).
19   Cong, L. & Singh, R. Symmetry-Protected Dual Bound States in the Continuum in Metamaterials. *Adv. Opt. Mater.* **7**, 1900383, (2019).
20   Liu, M. & Choi, D.-Y. Extreme Huygens' Metasurfaces Based on Quasi-Bound States in the Continuum. *Nano Lett.* **18**, 8062-8069, (2018).
21   Fan, K., Shadrivov, I. V. & Padilla, W. J. Dynamic bound states in the continuum. *Optica* **6**, 169-173, (2019).
22   Han, S. *et al.* All-Dielectric Active Terahertz Photonics Driven by Bound States in the Continuum. *Adv. Mater.* **31**, 1901921, (2019).
23   Li, Z. *et al.* Terahertz bound state in the continuum in dielectric membrane metasurfaces. *New J. Phys* **24**, 053010, (2022).
24   Rybin, M. V. *et al.* High-Q Supercavity Modes in Subwavelength Dielectric Resonators. *Phys. Rev. Lett.* **119**, 243901, (2017).
25   Kyaw, C. *et al.* Polarization-selective modulation of supercavity resonances originating from bound states in the continuum. *Commun. Phys.* **3**, 212, (2020).
26   Yu, N. *et al.* Light Propagation with Phase Discontinuities: Generalized Laws of Reflection and Refraction. *Science* **334**, 333-337, (2011).
27   Yang, Q. *et al.* Broadband terahertz rotator with an all-dielectric metasurface. *Photon. Res.* **6**, 1056-1061, (2018).
28   Yang, Q. *et al.* Mie-Resonant Membrane Huygens' Metasurfaces. *Adv. Func. Mater.* **30**, 1906851, (2020).
29   Quanlong, Y. *et al.* Topology-empowered membrane devices for terahertz photonics. *Adv. Photon.* **4**, 046002, (2022).
30   Romano, S. *et al.* Surface-Enhanced Raman and Fluorescence Spectroscopy with an All-Dielectric Metasurface. *J. Phys. Chem. C* **122**, 19738-19745, (2018).
31   Weber, T. *et al.* Intrinsic strong light-matter coupling with self-hybridized bound states in the continuum in van der Waals metasurfaces. *Nat. Mater.* **22**, 970-976, (2023).
32   Hou, L. *et al.* Probing trace lactose from aqueous solutions by terahertz time-domain spectroscopy. *Spectrochim. Acta, Part A* **246**, 119044, (2021).
33   Hou, L., Wang, J.-N., Wang, L. & Shi, W. Experimental study and simulation analysis of terahertz absorption spectra of α-lactose aqueous solution. *Acta Phys Sin-CH ED* **70**, 243202-243201-243202-243207, (2021).
34   Kikkawa, R., Nishida, M. & Kadoya, Y. Polarization-based branch selection of bound states in the continuum in dielectric waveguide modes anti-crossed by a metal grating. *New J. Phys* **21**, 113020, (2019).
35   Zhang, X. *et al.* Terahertz metasurface with multiple BICs/QBICs based on a split ring resonator. *Opt. Express* **30**, 29088-29098, (2022).
36   Grahn, P., Shevchenko, A. & Kaivola, M. Electromagnetic multipole theory for optical nanomaterials. *New J. Phys* **14**, 093033, (2012).

37   Friedrich, H. & Wintgen, D. Interfering resonances and bound states in the continuum. *Phys. Rev. A* **32**, 3231-3242, (1985).